\documentstyle[epsfig,psfig]{texas}

\def\Journal#1#2#3#4{{#1} {\bf #2}, #3 (#4)}

\def\NPA{{\em Nucl. Phys.} A}
\def\PLB{{\em Phys. Lett.}  B}
\def\PRL{\em Phys. Rev. Lett.}
\def\PRC{{\em Phys. Rev.} C}
\def\PRD{{\em Phys. Rev.} D}

\def\PPNP{\em Progr. Part. Nucl. Phys.}
\def\PR{\em Phys. Rept.}
\def\APJ{\em ApJ.}

\def\be{\begin{equation}}
\def\ee{\end{equation}}
\def\bea{\begin{eqnarray}}
\def\eea{\end{eqnarray}}

\def\bq{\begin{eqnarray}}
\def\eq{\end{eqnarray}}

\begin{document}

\title{NEUTRON STARS, SUPERNOVA EXPLOSIONS AND THE TRANSITION TO 
QUARK MATTER}
\author{Alessandro DRAGO}
\address{Dipartimento di Fisica - Universit\`a di Ferrara\\
Via Paradiso 12, 44100 FERRARA - ITALY\\
{\rm Email: drago@fe.infn.it}}

\begin{abstract}
The transition to quark matter can take place 
in neutron stars. 
The structure of a hybrid star, containing a core made
of quark matter is discussed. 
The maximum mass of the non-rotating hybrid star turns out to be
$\sim 1.6M_\odot$. Possible signatures of the 
quark phase as pulsar's timing, cooling rate etc. are briefly analyzed. 
The deconfinement transition can also take place during the 
pre-supernova collapse. This possibility is studied by introducing 
a finite temperature EOS.
The dependence of the latter on the proton fraction is shown to be crucial.
The softening of the EOS at densities just
above nuclear matter saturation density 
for $Z/A\sim 0.3$ helps in obtaining an explosion.
At the same time, at larger densities the EOS
is stiff enough to support a neutron star compatible with observations.
\end{abstract}

\section{Introduction}

The major problem in the study of the transition to quark matter
is the uncertainty in the choice of the relevant degrees of freedom
at the densities and temperatures at which the transition is supposed to take 
place. In the present contribution I will restrict the discussion
to the low temperatures regime, {\it i.e.} temperatures of the order
of few tens MeV. In this regime, the transition to quark matter will
be driven essentially by the density of the system, while the temperature
will play a relatively minor role, shifting the critical densities
to lower values.

The reasons why I will concentrate into the low temperature regime
are the following: a) neutron stars (NS), except during the very first
seconds of their life, are at temperatures of the order of few MeV or lower;
b) during the pre-supernova collapse, temperatures not exceeding few
tens MeV can be reached; finally c) quark models have been extensively
studied at low temperatures, where the experimental value 
of several observables constrains the value of the models' parameters.
In this contribution I will try to show that, at least in the above
described regime, a coherent scenario can be outlined. 

\section{Quark models}

Most of the calculations of the transition to quark matter are based
on the MIT bag model. 
The crucial parameter in this model is the so-called {\it pressure
of the vacuum} $B$, namely the energy necessary to dig a hole in the
non-perturbative QCD vacuum. Here are a few estimates of $B$:

\begin{itemize}
\item

from the computation of the hadronic spectrum \cite{degrand}

$B=59$ MeV/fm$^3$ or $B^{1/4}=145$ MeV

\item

from the computation of hadronic structure functions \cite{thomas}

$B\sim 109$ MeV/fm$^3$ or $B^{1/4} \sim 170$ MeV

\item

from the comparison of MIT bag pure-gauge results 
with lattice QCD at finite temperature 
\cite{satz}: a critical temperature $T_c= 240\pm20$ MeV for pure gauge SU(3)
corresponds to 

$B=170 \pm 50 $ MeV/fm$^3$ or $B^{1/4}= 190\pm 15$ MeV
\end{itemize}
The conclusions one can obtain from the previous estimates are the 
following: a) there is a certain ambiguity among the various calculations
of {\it single hadron} properties. The preferred value for $B$ is in the range
60 MeV/fm$^3 \le B \le $ 110 MeV/fm$^3$; b) the estimates based
on calculations trying to reproduce lattice QCD results indicate
a larger value for $B$. It is anyway important to stress that
at the moment there is no consensus about the relevant
degrees of freedom at finite temperature. For instance, gluons can develop
a thermal mass: if the latter is taken into account the lattice QCD
data can be reproduced using a very small value for $B$ \cite{greiner}.
It is also important to stress that for B$\ge$ 65 MeV/fm$^3$
iron is the ground state of matter. Therefore 
one is not obliged to accept Witten's strange matter hypothesis even using
for B a value indicated by single hadron calculations.

Among the many studies of the structure of NS based on
the MIT bag model, I would like to comment the most recent ones.
Akmal {\it et al.} \cite{akmal} 
have used, for the hadronic sector, a non-relativistic
Equation Of State (EOS) based on the Argonne potential and incorporating
three-body forces. In their analysis they consider two values for $B$:
122 and 200 MeV/fm$^3$. Using the smaller value the transition
to quark matter starts at $\rho_c\sim 3.5 \rho_0$ and the bulk of a heavy 
NS is made of a mixed phase of hadronic and quark matter. Clearly,
if a smaller value for $B$ would have been used, the mixed phase
would be present also in lighter NS. 

A similar calculation has been performed by IIda and Sato \cite{iida}.
In this case in the hadronic sector it has been used
a relativistic EOS which takes into account also hyperonic
degrees of freedom. The main result they obtain is that 
heavy NS contain quark matter essentially for all
values of $B$. Moreover, if $B$ is smaller than $\sim$
90 MeV/fm$^3$ quark matter is present even in `standard' NS
having a mass $M\sim$1.4 M$_\odot$.

The calculations that will be presented in the remaining part of 
this contribution 
are not based on the MIT bag model, but on
a non-topological soliton model called Color Dielectric Model (CDM) 
\cite{pirner,birse,banerjee}. 
The main differences respect to the MIT bag model are the following: 

\begin{itemize}
\item
MIT
\begin{itemize}
\item
current masses for the quarks
\item
large value of the pressure of the vacuum
\item
rigid confinement in a sphere
\item
ambiguous treatment of center of mass motion
\end{itemize}
\item
CDM
\begin{itemize}
\item
constituent masses for the quarks
\item
small value of the pressure of the vacuum
\item
soft confinement {\it via} interaction with a scalar field
\item
center of mass motion removed in the non-relativistic limit
\end{itemize}
\end{itemize}
In the past it has been shown that in the CDM,
using a fixed set of parameters' values
it is possible to study nucleon form factors, structure functions and
to reproduce the main features of low energy spectroscopy.
This same set of parameters' values has been used to study the transition
to quark matter.

\section{Neutron stars}
It has been investigated the structure of a {\it hybrid} star, which 
in the outer region is made of nucleonic matter and in the center
contains quark matter \cite{plb}. 

\subsection{Composition}
In most of the calculations a naive Walecka-type relativistic model
has been used to describe the hadronic phase.
Considering a NS having a mass $M=1.4 M_\odot$, half of the volume
is occupied by pure quark matter. The mixed phase extends over 1.5 Km.
The radius of the star is slightly larger than 10 Km.
The maximum mass for a non-rotating star is $M_{max}=1.59 M_\odot$.
Looking to the mass-radius relation it appears that the star is
more compact than a NS made only of nucleons. This can be relevant
in the light of recent estimates of the radii of NS \cite{bombaci}. 

It is worth mentioning the recently discovered possibility of
having, during the slow-down of millisecond pulsars,
an effect similar to the back-bending in nuclear
physics \cite{glendenning}. Since
a millisecond pulsar is strongly deformed, its central density
is reduced respect to a non-rotating star. It is therefore possible
that quark matter is formed during the slow-down of the pulsar.
Since quark matter is denser than normal matter, the moment of inertia
of the star reduces and the pulsar re-accelerates to conserve
the angular momentum. If detected, this anomalous behaviour 
of a millisecond pulsar
would be the signal that a deep modification in the composition
of the star is taking place. 

\subsection{Cooling}
The problem of the cooling of a NS has become very unclear in the last 
years \cite{page}.
It has been pointed out recently 
\cite{chabrier} that the relation between the
internal temperature of the NS, which depends on the composition
of the interior of the star, and the external temperature, that can be
measured using X-ray satellites, can be deeply modified by the
presence of light nuclei on the surface of the star.
The presence of light nuclei increases the heat transport and therefore the 
ratio between the internal and the external temperature could be
smaller than the previously estimated factor one hundred.

It is anyway interesting to remark that, using the CDM, the direct
URCA mechanism at the level of the quarks can take place, but
is strongly suppressed due to the extreme smallness of the
electron fraction in the interior of the NS. The computed cooling rate 
is only slightly faster than the modified URCA mechanism taking place
in a traditional NS made of nucleons. Therefore,
at variance with the MIT bag results, using the CDM is not necessary
to invoke any re-heating mechanism to obtain a cooling rate of the 
right order of magnitude.

\section{Supernova explosion}

The problem of getting a successful supernova explosion
is still open after many years of work. I will concentrate here
on the so-called direct mechanism. It is characterized by the idea 
that the explosion is directly related to the shock wave generated
by the bounce. During the last years this mechanism drop out of fashion,
and most of the recent studies are related to the idea of a neutrino
driven explosion, in which the initial shock stalls and the
explosion is revitalized by the energy deposited by the neutrinos
in the regions outside the forming proto-neutron star.

\begin{figure}

\centerline{\hbox{
\psfig{figure=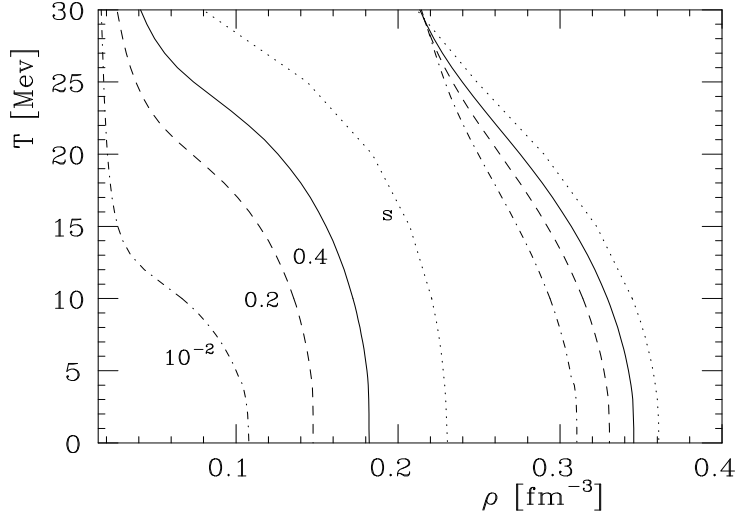,angle=90,height=7truecm,width=10truecm}
}}
\caption{Boundaries separating hadronic matter from mixed phase
(left lines) and the latter from quark matter (right lines).
The labels indicate various values of $Y_{l_e}$, $s$ is for symmetric matter. }
\end{figure}

The possibility of a
successful explosion via the direct mechanism is related to the
softness of the EOS at densities just above nuclear matter saturation density 
\cite{swesty}.
This possibility has been
ruled out for the only reason that seems to be inconsistent with the
1.44 $M_\odot$ constraint coming from PSR 1913+16 \cite{arnett}.
On the other hand Cooperstein concluded that if a phase transition
takes place during the collapse,
`the presence of a mixed-phase region
softens the EOS and leads to a direct explosion' \cite{cooperstein}.

The crucial problem to be analyzed is the dependence of the
critical densities on the proton fraction $Z/A$. Clearly no quark
matter has to be present in (nearly) symmetric nuclear matter
at densities of the order of the saturation density $\rho_0$. 
On the other hand,
it is exactly in this range of densities that the transition
should take place, when $Z/A\sim 0.3$, to influence the collapse
of the supernova. We have therefore explored the dependence
of the critical densities on the proton fraction \cite{astro}.
It is also important to check the effect of the temperature
on the transition. 

In Fig. 1 are presented
the boundaries separating hadronic matter from
mixed phase and the latter from pure quark matter.
The labels correspond to various values of the lepton fraction $Y_{l_e}$.
Symmetric nuclear matter is also presented.
The transition region depends on the lepton fraction $Y_{l_e}$.
In symmetric matter at low temperatures the mixed phase forms at
$\rho=0.23$ fm$^{-3}$, 
therefore no quark matter is present in heavy nuclei.
Decreasing the value of $Y_{l_e}$  
the phase transition starts at lower densities.
At any value of $Y_{l_e}$
the mixed phase extends on a rather limited range of densities
and even at zero temperature
pure quark matter phase is reached 
at densities slightly larger than $2 \rho_0$.
At higher temperatures the transition starts at lower densities. 

\begin{figure}

\centerline{\hbox{
\psfig{figure=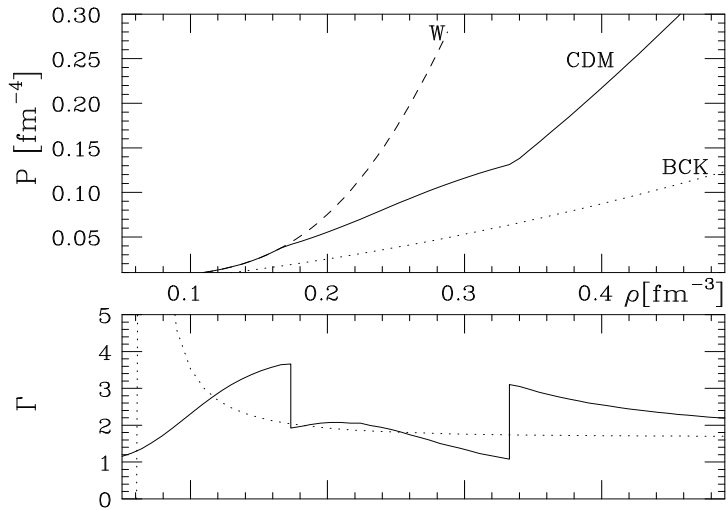,angle=90,height=7truecm,width=10truecm}
}}
\caption{Pressure (upper box) and adiabatic index (lower box) as function
of the density in CDM (solid) and in BCK (dotted). The pressure in Walecka
model is also shown (dashed). }
\end{figure}

To investigate our EOS in connection with the problem of supernova
explosion, we compare with BCK EOS \cite{bck}. 
The latter is a totally phenomenological EOS which is soft enough
to allow for supernova explosion, but gives a maximum mass smaller
than the mass of PSR 1913+16.

In Fig. 2 are presented
results for $Y_{l_e}=0.4$ and entropy per baryon number
$S/R=1$.  
In the upper box we
compare the pressure in the Walecka model, in our model and in BCK
EOS.  Due to the phase transition, our EOS is rather soft from $\rho=0.17$
fm$^{-3}$ to $\rho=0.34$ fm$^{-3}$.  On the other hand, after $\rho =
0.34$ fm$^{-3}$ it is considerably stiffer than BCK, allowing higher masses
for the proto-neutron star. These conclusions are strengthened by the
computation of the adiabatic index, shown in the lower box of 
Fig. 2.
Clearly in the mixed phase matter offers little resistance to
collapse, but when pure quark matter phase is reached the collapse is
halted. In the mixed phase region our adiabatic index is even smaller
than in BCK.
It is important to remark that similar results have been obtained
using the Lattimer-Swesty EOS \cite{ls} to describe the hadronic phase.

\section{Conclusions}

The main results are the following.
Calculations based on various quark models indicate the
presence of quark matter at least in the center of heavy NS.
Moreover, if the quark model's 
parameters are fixed to reproduce basic hadronic properties then
a) the transition to quark matter takes place at low densities in
$\beta$-stable matter and
b) neutron stars having a mass $M \sim 1.4 M_\odot$ contain
quark matter. Hybrid stars are 
characterized by a small $R/M$ ratio.

A signal of a transition to quark matter would be  
the spin-up of a millisend pulsar, due to the modification of its
moment of inertia. 

Finally,
the transition to quark matter could take place during heavy star
collapse, helping supernova to explode.

\end{document}